\documentstyle[twoside]{article}
\input{psfig}

\newcommand{\lesssim}{\lower 2pt \hbox{$\:\stackrel{<}{\scriptstyle \sim}\:$}}
\newcommand{\gtrsim}{\lower 2pt \hbox{$\:\stackrel{>}{\scriptstyle \sim}\:$}}
\newcommand{\case}[2]{\textstyle \frac{#1}{#2}}

\newcounter{dum}
\newenvironment{mathletters}
{\refstepcounter{equation}\setcounter{dum}{\value{equation}}%
 \setcounter{equation}{0}%
 }
{\setcounter{equation}{\value{dum}}}

\begin{document}

\thispagestyle{empty}

\begin{center}
{\Large\bf Shot noise in mesoscopic systems}\\
\bigskip
{M. J. M. de Jong}\\
\smallskip
{\em Philips Research Laboratories\\ 5656 AA Eindhoven, The Netherlands}\\
\medskip
{and}\\
\medskip
{C. W. J. Beenakker}\\
\smallskip
{\em Instituut--Lorentz, University of Leiden\\
2300 RA Leiden, The Netherlands}
\end{center}

\begin{abstract}
This is a review of shot noise, the time-dependent fluctuations in the
electrical current due to the discreteness of the electron charge,
in small conductors.
The shot-noise power can be smaller than that of a Poisson process
as a result of correlations in the electron transmission
imposed by the Pauli principle.
This suppression takes on simple universal values in a symmetric
double-barrier junction (suppression factor $\case{1}{2}$), a disordered
metal (factor $\case{1}{3}$), and a chaotic cavity
(factor $\case{1}{4}$). Loss of phase coherence
has no effect on this shot-noise suppression, while thermalization of
the electrons due to electron-electron scattering increases the
shot noise slightly. Sub-Poissonian shot noise has been observed
experimentally. So far unobserved phenomena involve the interplay
of shot noise with the Aharonov-Bohm effect, Andreev reflection,
and the fractional quantum Hall effect.
\end{abstract}

\section{Introduction}
\label{q1}

\subsection{Current fluctuations}
\label{q1A}

In 1918 Schottky \cite{sch18} reported that in ideal vacuum tubes, where all
sources of spurious noise had been eliminated, there remained two types of
noise in the electrical current, 
described by him as the {\em W\"{a}rmeeffekt\/} and the {\em
Schroteffekt}. The first type of noise became known as Johnson-Nyquist noise
(after the experimentalist \cite{joh27} and the theorist \cite{nyq28} who
investigated it), or simply thermal noise. It is due to the thermal motion
of the electrons and occurs in any conductor. The second type of noise is
called shot noise, caused by the discreteness of the charge of the carriers of
the electrical current. Not all conductors exhibit shot noise.

Noise is characterized by its spectral density 
or power spectrum $P(\omega)$, which is
the Fourier transform at frequency $\omega$
of the current-current correlation function \cite{Ziel,Dav},
\begin{equation}
P(\omega) = 2 \int \limits_{-\infty}^\infty dt \, e^{i \omega t} 
\langle \Delta I(t+t_0) \Delta I(t_0) \rangle
\: .
\label{q1B.1}
\end{equation}
Here $\Delta I(t)$ denotes the time-dependent fluctuations
in the current at a given voltage $V$ and temperature ${\rm T}$. 
The brackets $\langle \cdots \rangle$ indicate an ensemble average
or, equivalently, an average over the initial time $t_0$.
Both
thermal and shot noise have a white power spectrum --- that is, the noise power
does not depend on $\omega$ over a very wide frequency range. Thermal noise
($V=0, \: {\rm T}\neq 0$) is directly related to the conductance $G$ by the
fluctuation-dissipation theorem \cite{cal51},
\begin{equation}
P = 4 k_B {\rm T} G \: ,
\label{q1B.2}
\end{equation}
as long as $\hbar \omega \ll k_B {\rm T}$.
Therefore,
the thermal noise of a conductor does not give any new information.

Shot noise ($V\neq 0, \:  {\rm T}=0$) is more interesting, because it gives information
on the temporal correlation of the electrons, which is not contained in the
conductance. In devices such as tunnel junctions, Schottky barrier diodes,
$p$-$n$ junctions, and thermionic vacuum diodes \cite{Ziel},
the electrons are transmitted
randomly and independently of each other. The transfer of electrons can be
described by Poisson statistics, which is used to analyze events that are
uncorrelated in time. For these devices the shot noise has its maximum value
\begin{equation}
P=2 e I \equiv P_{\rm Poisson} \: ,
\label{q1B.4}
\end{equation}
proportional to the time-averaged current $I$. 
(We assume $I>0$ and $V>0$ throughout this review.)
Equation (\ref{q1B.4}) is valid for $\omega < \tau^{-1}$,
with $\tau$ the width of a one-electron current pulse.
For higher frequencies the shot noise vanishes.
Correlations suppress the low-frequency
shot noise below $P_{\rm Poisson}$. One source of
correlations, operative even for non-interacting electrons, is the Pauli
principle, which forbids multiple occupancy of the same single-particle state.
A typical example is a ballistic point contact in a metal, where $P=0$ because
the stream of electrons is completely correlated by the Pauli principle in the
absence of impurity scattering. Macroscopic, metallic conductors have zero shot
noise for a different reason, namely that inelastic electron-phonon scattering
averages out the current fluctuations.

Progress in nanofabrication technology has revived the interest in shot
noise, because nanostructures allow measurements to be made on
``mesoscopic'' length scales that were previously inaccessible. The
mesoscopic length scale is much greater than atomic dimensions, but
small compared to the scattering lengths associated with various
inelastic processes. Mesoscopic systems have been studied extensively
through their conductance \cite{Imry,C&H,MPiS,LesH}. Noise measurements are
much more difficult, but the sensitivity of the experiments has made a
remarkable progress in the last years. Some theoretical predictions have
been observed, while others still remain an experimental challenge.
This article is a review of the present status of the field, with an
emphasis on the theoretical developments. We will focus on the
scattering approach to electron transport, which provides a unified
description of both conductance and shot noise. For earlier reviews,
see Refs.\ \cite{mar94,lan95,but96}.
For brief commentaries, see Refs.\ \cite{kou96,jon96d}.

\subsection{Scattering Theory}
\label{q1D}

In his 1957 paper \cite{lan57} Landauer discussed the problem of 
electrical conduction as a scattering problem.
This has become a key concept in mesoscopic
physics \cite{Imry,C&H}.
The conductor is modeled as a scattering region, connected
to electron reservoirs. The electrons inside each reservoir are assumed to
be in thermal equilibrium.
Incoming states, occupied according to the Fermi-Dirac
distribution function, are scattered into outgoing states.
At low temperatures the conductance is fully determined
by the transmission matrix of electrons at the Fermi level. 
The two-terminal Landauer formula \cite{Imry,fis81} and 
its multi-terminal generalization \cite{but86,sto88,bar89,she91}
constitute a general framework
for the calculation of the conductance of a phase-coherent sample.
A scattering theory of the noise properties of 
mesoscopic 
conductors was derived in Refs.\ \cite{khl87,lan89,les89,yur90,but90,lan91,but92b}. 
The basic result is a relationship between the shot-noise power and the
transmission matrix at the Fermi level, analogous to the Landauer formula
for the conductance. Here we review the derivation of this result,
following closely B\"{u}ttiker's work \cite{but90,but92b}.

\begin{figure}[tb]
\centerline{\psfig{file=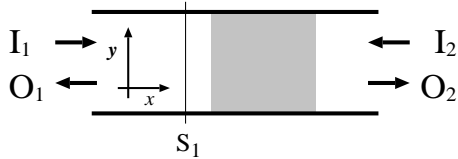,width=6cm}}
\caption{
Schematic representation of the transport through the conductor.
Incoming states ($I$) are scattered into outgoing states ($O$),
by a scattering region (dashed).
A cross section in lead 1 and
its coordinates are indicated. 
}
\label{q1D.f2}
\end{figure}

Two leads are connected to an arbitrary 
scattering region (see Fig.\ \ref{q1D.f2}).
Each lead contains $N$ incoming and $N$ outgoing
modes at energy $\varepsilon$.
We assume only elastic scattering
so that energy is conserved.
The incoming and outgoing modes are related by a
$2 N \times 2 N$ scattering matrix ${\sf S}$
\begin{equation}
\left( \begin{array}{c} O_1 \\ O_2 \end{array} \right) =
{\sf S} 
\left( \begin{array}{c} I_1 \\ I_2 \end{array} \right)
\: ,
\label{q1D.5}
\end{equation}
where $I_1,O_1,I_2,O_2$ are the $N$-component vectors denoting the
amplitudes of
the incoming ($I$) and outgoing ($O$) modes in lead 1 and lead 2.
The scattering matrix can be decomposed in
$N \times N$ reflection and transmission matrices,
\begin{equation}
{\sf S}=\left( 
\begin{array}{cc} {\sf s}_{11} & {\sf s}_{12} \\ {\sf s}_{21}  & {\sf
s}_{22}
\end{array} \right)
\equiv \left( 
\begin{array}{cc} {\sf r} & {\sf t}' \\ {\sf t}  & {\sf r}'
\end{array} \right)
\: ,
\label{q1D.4}
\end{equation}
where the $N\times N$ matrix ${\sf s}_{ba}$ 
contains the amplitudes $s_{bn,am}$
from incoming mode $m$ in lead $a$ to
outgoing mode $n$ in lead $b$.
Because of flux conservation $\sf S$ is a unitary matrix.
Moreover, in the presence of time-reversal symmetry $\sf S$ is symmetric.

The current operator in lead 1 is given by
\begin{equation}
\hat{I}(t) = \frac{e}{h} \sum_{\alpha,\beta}
\int \limits_0^\infty d \varepsilon 
\int \limits_0^\infty d \varepsilon' \,
I_{\alpha \beta}(\varepsilon,\varepsilon') \,
\hat{a}_\alpha^\dagger (\varepsilon) \,
\hat{a}_\beta^{\vphantom{\dagger}} (\varepsilon') \,
e^{i t (\varepsilon-\varepsilon')/\hbar }
\: ,
\label{q1D.7}
\end{equation}
where $\hat{a}_\alpha^\dagger (\varepsilon)$
[$\hat{a}_\alpha^{\vphantom{\dagger}} (\varepsilon)$] is the creation [annihilation]
operator of scattering state $\psi_\alpha({\bf r},\varepsilon)$.
We have introduced 
the indices $\alpha \equiv (a,m)$, $\beta \equiv (b,n)$
and the coordinate ${\bf r}=(x,{\bf y})$.
The matrix element $I_{\alpha \beta}(\varepsilon,\varepsilon')$
is determined by the value of the current at cross section $S_1$
in lead 1,
\begin{equation}
I_{\alpha \beta}(\varepsilon,\varepsilon') =
\case{1}{2} \int \limits_{S_1} \! \! d {\bf y} \! \left\{
\psi_\alpha({\bf r},\varepsilon) 
[ \hat{v}_x
\psi_\beta({\bf r},\varepsilon') ]^* +
\psi_\beta^*({\bf r},\varepsilon') 
\hat{v}_x \psi_\alpha({\bf r},\varepsilon) 
\right\}
\, .
\label{q1D.8}
\end{equation}
Here, $\hat{v}_x$ is the velocity operator in the $x$-direction.
At equal energies, Eq.\ (\ref{q1D.8}) simplifies to \cite{but90,but92b}
\begin{equation}
I_{am,bn}(\varepsilon,\varepsilon) =
\delta_{a1} \delta_{ab} \delta_{mn} - 
\sum_{p=1}^N s_{1p,am}(\varepsilon) \, s_{1p,bn}^*(\varepsilon)
\: .
\label{q1D.9}
\end{equation}

The average current follows from 
\begin{equation}
\langle  \hat{a}_\alpha^\dagger (\varepsilon)
\hat{a}_\beta^{\vphantom{\dagger}} (\varepsilon') \rangle
=\delta_{\alpha \beta} \delta(\varepsilon -\varepsilon')
f_a(\varepsilon)
\: ,
\label{q1D.10}
\end{equation}
where $f_a$ is the Fermi-Dirac distribution function in reservoir $a$:
\begin{mathletters}
\label{q1D.11}
\begin{eqnarray}
f_1(\varepsilon) &=& f(\varepsilon - E_F - eV) \: ,
\\
f_2(\varepsilon) &=& f(\varepsilon - E_F ) \: , \\
f(x)&=&[1+\exp(x/k_B {\rm T})]^{-1}\: ,
\end{eqnarray}
\end{mathletters}%
with Fermi energy $E_F$.
The result is 
\begin{equation}
\langle \hat{I}(t) \rangle = \frac{e}{h} 
\sum_\alpha \int \limits_0^\infty \! \!
d\varepsilon f_a(\varepsilon)
I_{\alpha \alpha}(\varepsilon,\varepsilon) 
= \frac{e}{h} \int \limits_0^\infty \! \!
d\varepsilon \left[ f_1(\varepsilon) 
- f_2(\varepsilon) \right]
\, \mbox{Tr} \, 
{\sf t}(\varepsilon) {\sf t}^\dagger(\varepsilon) \: ,
\label{q1D.12}
\end{equation}
where we have substituted Eq.\ (\ref{q1D.9}) 
and used   
the unitarity of $\sf S$. The linear-response conductance,
$G \equiv \lim_{V \rightarrow 0} \langle I \rangle / V$,
becomes
\begin{equation}
G= \frac{e^2}{h} \int \limits_0^\infty
d\varepsilon \left( - \frac{\partial f}{\partial \varepsilon} \right)
\mbox{Tr} \, {\sf t}(\varepsilon) {\sf t}^\dagger(\varepsilon) 
\: ,
\label{q1D.13}
\end{equation}
which at zero temperature simplifies to the Landauer formula
\begin{equation}
G = \frac{e^2}{h} \mbox{Tr}\, {\sf t} \, {\sf t}^\dagger =
\frac{e^2}{h} \sum_{n=1}^{N} T_n \: .
\label{q1D.14}
\end{equation}
Here ${\sf t}$ is taken at $E_F$ and $T_n\in [0,1]$ is an eigenvalue of
${\sf t} \, {\sf t}^\dagger$. The
conductance is thus fully
determined by the transmission eigenvalues.
Knowledge of the 
transmission eigenstates, each of which can be a complicated superposition of
incoming modes, is not required.

In order to evaluate the shot-noise power we 
substitute the current operator (\ref{q1D.7})
into Eq.\ (\ref{q1B.1}) and determine
the expectation value.
We use the formula \cite{but92b}
\begin{equation}
\langle \hat{a}_1^\dagger \hat{a}_2^{\vphantom{\dagger}} \hat{a}_3^\dagger
\hat{a}_4^{\vphantom{\dagger}} 
\rangle -
\langle \hat{a}_1^\dagger \hat{a}_2^{\vphantom{\dagger}}\rangle
\langle \hat{a}_3^\dagger \hat{a}_4^{\vphantom{\dagger}} 
\rangle
=  \delta_{14} \delta_{23} f_1 (1-f_2) \equiv \Delta_{1234}  \: ,
\label{q1D.15}
\end{equation}
where {\em e.g.}\ $\delta_{12}$ stands for 
$\delta_{\alpha\beta} \delta(\varepsilon-\varepsilon')$.
Equation (\ref{q1D.15})
shows that there are cross correlations between different
scattering states.
Although this bears no effect on the time-averaged current,
it is essential for the current fluctuations.
For the noise power one finds
\begin{eqnarray}
P(\omega)&=&2 \, \frac{e^2}{h^2} \sum_{\alpha,\beta,\gamma,\delta}
\; \int \limits_{-\infty}^\infty dt
\int \limits_0^\infty d\varepsilon 
\int \limits_0^\infty d\varepsilon'
\int \limits_0^\infty d\varepsilon''
\int \limits_0^\infty d\varepsilon''' e^{i(\hbar \omega +
\varepsilon-\varepsilon')t/\hbar}
\nonumber \\
&\times&
I_{\alpha\beta} (\varepsilon,\varepsilon') 
I_{\gamma\delta} (\varepsilon'',\varepsilon''')
\Delta_{\alpha\beta\gamma\delta} 
(\varepsilon,\varepsilon',\varepsilon'',\varepsilon''')
\nonumber \\
&=&2 \, \frac{e^2}{h} \sum_{\alpha,\beta} 
\int \limits_0^\infty d\varepsilon 
I_{\alpha\beta} (\varepsilon,\varepsilon+\hbar \omega ) 
I_{\beta\alpha} (\varepsilon + \hbar \omega,\varepsilon)
f_a(\varepsilon) [ 1 - f_b(\varepsilon+\hbar \omega)] \: .
\nonumber \\
&&
\label{q1D.17}
\end{eqnarray}
The low-frequency limit is found by substitution of 
Eq.\ (\ref{q1D.9}),
\begin{eqnarray}
P&=&2 \, \frac{e^2}{h} \int \limits_0^\infty d\varepsilon 
\left\{ 
[  f_1 (1-f_2) + 
f_2(1-f_1) ]
\, \mbox{Tr} \, {\sf t\, t}^\dagger 
( {\sf 1} - {\sf t \, t}^\dagger ) 
\right. \nonumber \\
&& + \,
\left.
[ f_1(1-f_1)
+ f_2(1-f_2)]
\, \mbox{Tr} \, {\sf t \, t}^\dagger {\sf t \, t}^\dagger 
\right\} \: ,
\label{q1D.18}
\end{eqnarray}
where we have again used the unitarity of $\sf S$.

Equation (\ref{q1D.18}) allows us to evaluate the noise for various cases.
Below we will assume that $eV$ and $k_B$T are small enough to 
neglect the energy dependence of the transmission matrix, so that we
can take ${\sf t}$ at $\varepsilon=E_F$.
Let us first determine the noise in equilibrium, {\em i.e.}\ for $V=0$.
Using the relation $f(1-f)=-k_B$T$\partial f/\partial \varepsilon$ we find
\begin{equation}
P=4 k_B \mbox{T} \frac{e^2}{h} \mbox{Tr} \, {\sf t \, t}^\dagger
= 4 k_B \mbox{T} \frac{e^2}{h}
\sum_{n=1}^{N} T_n \: ,
\label{q1D.19}
\end{equation}
which is indeed the Johnson-Nyquist formula (\ref{q1B.2}).
For the shot-noise power at zero temperature we obtain
\begin{equation}
P = 2 e V \frac{e^2}{h} \, \mbox{Tr} \,
{\sf t} \, {\sf t}^\dagger
( {\sf 1} - {\sf t} \,
{\sf t}^\dagger ) = 2 e V \frac{e^2}{h} \, \sum_{n=1}^{N}
T_n ( 1 - T_n ) \: .
\label{q1D.20}
\end{equation}
Equation (\ref{q1D.20}), due to B\"{u}ttiker \cite{but90},
is the multi-channel generalization of the
single-channel formulas of Khlus \cite{khl87},
Lesovik \cite{les89}, and Yurke and Kochanski \cite{yur90}.
One notes, that $P$ is again only a function of the transmission
eigenvalues.

It is clear from Eq.\ (\ref{q1D.20}) that a transmission eigenstate for
which $T_n=1$ does not contribute to the shot noise. 
This is easily understood:
At zero temperature there is a non-fluctuating incoming electron 
stream. If there is complete transmission, the transmitted electron
stream will be noise free, too.
If $T_n$ decreases, the transmitted electron stream deviates in time
from the average current. The resulting shot noise $P$ is still smaller
than $P_{\rm Poisson}$, because the transmitted electrons are correlated due
to the Pauli principle. Only if $T_n \ll 1$, the transmitted electrons are
uncorrelated, yielding full Poisson noise (see Sec.\ \ref{q1Ea}).
Essentially, the non-fluctuating occupation number of the incoming states
is a consequence of the electrons being fermions. In this sense, the
suppression below the Poisson noise is due to the Pauli principle.
On the other hand, one must realize that
the noise suppression is not an exclusive property of fermions.
It occurs for any incoming beam with a non-fluctuating occupation number,
for example a photon number state \cite{liu95}.

The generalization of Eq.\ (\ref{q1D.20}) to the non-zero voltage, non-zero
temperature case is \cite{but92b,mar92}
\begin{equation}
P=2 \, \frac{e^2}{h} \sum_{n=1}^N \left[ 2 k_B \mbox{T} \, T_n^2 +
T_n(1-T_n) eV \coth (eV/2k_B \mbox{T})  \right] \: .
\label{q1D.21}
\end{equation}
The crossover from the thermal noise (\ref{q1D.19}) to the shot noise
(\ref{q1D.20})
depends on the transmission eigenvalues. 

As a final remark, we mention that in the above derivations
the absence of spin and valley
degeneracy has been assumed for notational convenience. 
It can be easily included. For a two-fold spin degeneracy 
this results in the replacement of the $e^2/h$ prefactors
[such as in Eqs.\ (\ref{q1D.14}) and (\ref{q1D.20})] by $2 e^2 / h$.
From now on, we simply use $G_0\equiv\,$degeneracy factor$\,\times \, e^2 / h$
as the unit of conductance and
$P_0\equiv 2 e V G_0$ as the unit of shot-noise power.

\subsection{Two Simple Applications}
\label{q1E}

The above results are valid for conductors with arbitrary
(elastic) scattering. If the transmission eigenvalues 
are known, the conduction and noise properties can be readily
calculated.
Below, this is illustrated for two simple systems.
More complicated conductors are discussed in 
Secs.\ \ref{q2}--\ref{q4}.

\subsubsection{Tunnel barrier}
\label{q1Ea}

In a tunnel barrier, electrons have a very small probability
of being transmitted. We model this by taking
$T_n \ll 1$, for all $n$.
Substitution into the formula for the
shot noise (\ref{q1D.20}) and
the Landauer formula for the conductance (\ref{q1D.14}) yields
$P = P_{\rm Poisson}$ at zero temperature.
For arbitrary temperature we obtain
from  Eq.\ (\ref{q1D.21}),
\begin{equation}
P= \coth(eV/2k_B \mbox{T}) \, P_{\rm Poisson} \: .
\label{q1E.1}
\end{equation}
This equation, due to Pucel \cite{puc61},
describes the crossover from thermal noise to
full Poisson noise.
For tunnel barriers this crossover is governed entirely by
the ratio $eV/k_B$T and not by details of the
conductor. This behavior has been observed in various
systems, see {\em e.g.}\ Refs.\ \cite{liu67,lec68}.
Electron-electron interactions can lead to modification
of Eq.\ (\ref{q1E.1}), see Ref.\ \cite{ho83,sch85,lee96}.

\subsubsection{Quantum Point Contact}
\label{q1Eb}

A point contact is a narrow constriction between two pieces of 
conductor. If the 
width $W$ of the constriction is much smaller than 
the mean free path of the bulk material, 
but much greater than the Fermi wave length $\lambda_F$,  
the conductance is 
given by the Sharvin formula \cite{sha65}, which in two dimensions
reads $G =G_0 2 W / \lambda_F$.
In such a {\em classical} point contact the shot noise
is absent, as found by Kulik and Omel'yanchuk \cite{kul84}.
In a {\em quantum} point contact $W$ is comparable
to $\lambda_F$.
Experimentally, a quantum point contact can be formed in a two-dimensional
electron gas in an (Al,Ga)As heterostructure \cite{C&H}.
The constriction is defined by depletion of the electron gas
underneath metal gates on top 
of the structure. Upon changing the gate voltage $V_g$,
the width $W$ is varied.
The conductance displays a stepwise
increase in units of $G_0$ as a function of $V_g$ \cite{wee88,wha88}.
This is caused by the discrete number $N_0={\rm Int}[2W / \lambda_F]$
of modes at the Fermi energy which 
fit into the constriction width.
As a result, $N_0$ transmission eigenvalues equal 1, the others 0,
yielding a quantized conductance according
to Eq.\ (\ref{q1D.14}).

\begin{figure}[tb]
\centerline{
\psfig{file=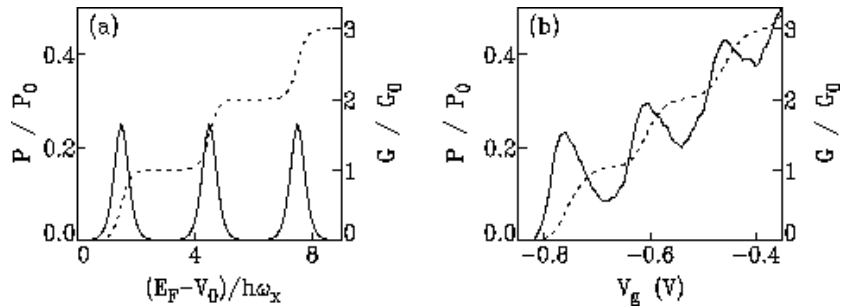,width=11cm}}
\caption{(a)
Conductance $G$ (dashed line) and shot-noise power $P$ (full line)
versus Fermi energy
of a two-dimensional quantum point contact, according to the 
saddle-point model, 
with $\omega_y = 3 \omega_x$.
(b) Experimentally observed $G$ and $P$ versus gate voltage $V_g$
(unpublished data from Reznikov {\em et al.} 
similar to the experiment of  Ref.\ \protect\cite{rez95}, but
at a lower temperature ${\rm T}= 0.4 \:$K).
}
\label{q1E.f1}
\end{figure}

Lesovik has predicted that the
shot noise in a quantum point contact is distinct from
a classical point contact \cite{les89}.
At the conductance plateaus the shot noise is absent,
as follows from Eq.\ (\ref{q1D.20}).   
However, in between the plateaus,
where the conductance increases by $G_0$, 
there is a transmission eigenvalue which is between 0 and 1.
As a consequence, the shot noise has a peak. 
We illustrate this behavior with a model by 
B\"{u}ttiker \cite{but90b}
of a two-dimensional saddle-point potential,
\begin{equation}
V(x,y) = V_0 - \case{1}{2} m \omega_x^2 x^2 + \case{1}{2} m
\omega_y^2 y^2
\: ,
\label{q1E.11}
\end{equation}
where $V_0$ is the potential at the saddle point, and $\omega_x$ and
$\omega_y$ determine the curvatures. 
The transmission eigenvalues at the Fermi energy 
are \cite{but90b}
\begin{equation}
T_n=[1+\exp(-2 \pi \varepsilon_n / \hbar \omega_x)]^{-1} \: ,
\;\;\;
\varepsilon_n \equiv E_F - V_0 - (n-\case{1}{2}) \hbar \omega_y \: .
\label{q1E.12}
\end{equation}
Results for the conductance and the shot-noise power 
are displayed in Fig.\ \ref{q1E.f1}a.
The shot noise peaks in between the conductance plateaus and is absent
on the plateaus.
For large $N$, the peaks in the shot noise become negligible
with respect to the Poisson noise, in agreement with
the classical result \cite{kul84}.
More theoretical work on noise in quantum point contacts is given in
Refs.\ \cite{yan92,kuh92,bee94,che95}. 

The prediction by Lesovik of this quantum size-effect in the
shot noise formed a challenge for experimentalists.
An early experiment was done by 
Li {\em et al.}\ \cite{li90a}. 
However, a difficulty in the interpretation was
that the frequency was not high enough to distinguish
between shot noise and
resistance fluctuations, which are also quite sensitive to
changes in the width of the point contact \cite{tim90,dek91}.
Recent experiments at much higher frequencies 
by Reznikov {\em et al.}\ \cite{rez95} (see also Ref.\ \cite{liu96})
and with very 
elaborate shielding by Kumar {\em et al.}\ \cite{kum96}
have unambiguously demonstrated the occurrence of suppressed shot noise
on the conductance plateaus.
Experimental data of Reznikov {\em et al.}\ are shown in 
Fig.\ \ref{q1E.f1}b.

\subsection{Kinetic Theory}
\label{q1F}

The scattering theory of Sec.\ \ref{q1D} fully takes into account the
phase coherence of the electron wave function. If phase coherence is
not essential, one can use instead a semiclassical kinetic theory. The
word ``semiclassical'' means that classical mechanics is combined with
the quantum-mechanical Pauli principle. A semiclassical kinetic theory
for shot noise has been developed by Kulik and Omel'yanchuk 
for a point contacts\cite{kul84},
by Nagaev for a diffusive conductor \cite{nag92},
and by the authors for an arbitrary conductor \cite{jon95c,jon96a}.
(Refs.\ \cite{jon95c,jon96a} correct Ref.\ \cite{bee91}.)

The theory is based on an extension of the
Boltzmann equation to include fluctuations of the distribution function
\cite{kad57,kog69,gan79}. By analogy with the Langevin equation in the
theory of stochastic processes, this fluctuating Boltzmann equation is
called the {\em Boltzmann-Langevin equation}. We give a brief summary
of the method.

The fluctuating distribution function
$f({\bf r}, {\bf k}, t)$ 
in the conductor
equals $(2 \pi)^d$ times the density
of electrons with position ${\bf r}$,
and wave vector ${\bf k}$, at time $t$.
The average over time-dependent fluctuations
$\langle f \rangle \equiv \bar{f}$
obeys the Boltzmann equation,
\begin{mathletters}
\label{q1F.1}
\begin{equation}
\left( \frac{d}{dt} + {\cal S} \right) 
\bar{f}({\bf r}, {\bf k}, t) = 0 \: ,
\label{q1F.1a}
\end{equation}
\begin{equation}
\frac{d}{dt} \equiv
\frac{\partial}{\partial t} + 
{\bf v} \cdot \frac{\partial}{\partial {\bf r}} + 
{\cal F} \cdot \frac{\partial}{\hbar \partial {\bf k}} \: .
\label{q1F.1b}
\end{equation}
\end{mathletters}%
The derivative (\ref{q1F.1b})
(with ${\bf v}=\hbar {\bf k}/m$)
describes the classical motion in the force field 
${\cal F}({\bf r})=-e \partial \phi({\bf r}) / \partial {\bf r}
+ e {\bf v} \times {\bf B}({\bf r})$,
with electrostatic potential $\phi({\bf r})$
and magnetic field ${\bf B}({\bf r})$.
The term ${\cal S} \bar{f}$ accounts for the
stochastic effects of scattering. 
In the case of impurity scattering,
the scattering term equals
\begin{equation}
{\cal S} f({\bf r},{\bf k},t)
= \int d {\bf k}' \, W_{ {\bf k}{\bf k}'}({\bf r}) [ 
f({\bf r},{\bf k},t)  - f({\bf r},{\bf k}',t) ]
\: .
\label{q1F.2}
\end{equation}
The kernel $W_{ {\bf k}{\bf k}'}({\bf r})$ is the transition rate 
for scattering from ${\bf k}$ to ${\bf k}'$, which may in
principle also depend on ${\bf r}$.

We consider the stationary situation, where
$\bar{f}$ is independent of $t$.
The time-dependent fluctuations
$\delta f \equiv f - \bar{f}$ satisfy
the Boltzmann-Langevin equation \cite{kad57,kog69},
\begin{equation}
\left( \frac{d}{dt} + {\cal S} \right)
\delta f ({\bf r}, {\bf k}, t) = 
j ({\bf r}, {\bf k}, t) \: ,
\label{q1F.3}
\end{equation}
where $j$ is a fluctuating source term
representing the fluctuations induced by the stochastic nature of the
scattering.
The flux $j$ has zero average,
$\langle j \rangle = 0$, and covariance 
\begin{equation}
\langle j ({\bf r}, {\bf k}, t) \, j ({\bf r}', {\bf k}', t') \rangle
= 
(2 \pi)^d \, \delta({\bf r} -{\bf r'}) \, \delta(t-t')
J({\bf r}, {\bf k}, {\bf k}') \: .
\label{q1F.4}
\end{equation}
The delta functions ensure that 
fluxes are only correlated
if they are induced by the same scattering process.
The flux correlator $J$ depends on the type of scattering and on $\bar{f}$, 
but not on $\delta f$.
Due to the Pauli principle the scattering possibilities of an incoming
state depend on the occupation of possible outgoing states.
As a consequence, $J$ is roughly proportional to
$\bar{f}(1-\bar{f}')$.
The precise correlator $J$ for the 
impurity-scattering term (\ref{q1F.2}) has been
derived by Kogan and Shul'man \cite{kog69}.
Scattering by a tunnel barrier corresponds to another correlator
\cite{jon95c,jon96a}.

The kinetic theory can be applied to calculate various
noise properties, including the effects of
electron-electron and 
electron-phonon scattering \cite{nag95,koz95,gur96}.
In Refs.\ \cite{jon95c,jon96a} a general formula for the shot-noise power
has been derived from Eqs.\ (\ref{q1F.3}) and (\ref{q1F.4}).
Further discussion of the kinetic theory is outside the scope of this
review.
In the following Section, we discuss an alternative method
to calculate the effects of phase breaking and other scattering processes.

\subsection{Phase Breaking, Thermalization,\\ and Inelastic Scattering}
\label{q1G}

Noise measurements require rather high currents, which
enhance the rate of scattering processes other than purely elastic
scattering.
The phase-coherent transmission approach of Sec.\ \ref{q1D} is then
no longer valid. 
The effects of dephasing and inelastic scattering on the
shot noise have been studied in Refs.\
\cite{nag92,jon96a,nag95,koz95,gur96,shi92,ued93,b&b92,lan93,liu94,zhe95}.
Below, we discuss a model \cite{jon96a,b&b92}
in which the conductor is divided in
separate, phase-coherent
parts connected by charge-conserving reservoirs.
This model includes
the following types of scattering:
\begin{itemize}
\item
{\em Quasi-elastic scattering.} Due to weak coupling with external
degrees of freedom
the electron-wave function gets dephased, but its energy is conserved. 
In metals, this scattering is caused by
fluctuations in the electromagnetic field  \cite{emf}.
\item
{\em Electron heating.}
Electron-electron scattering ex\-changes energy
between the electrons, but the total energy of the electron gas is conserved.
The distribution function is therefore assumed to be a 
Fermi-Dirac distribution at a temperature above the lattice temperature.
\item
{\em Inelastic scattering.} 
Due to electron-phonon interactions the
electrons exchange energy with the lattice. The electrons emerging from
the reservoir are distributed according to the Fermi-Dirac distribution,
at the lattice temperature ${\rm T}$.
\end{itemize}

\begin{figure}[tb]
\centerline{\psfig{file=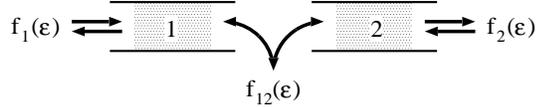,width=7cm}}
\caption{
Additional scattering inside the conductor is modeled
by dividing it in two parts and connecting them
through another reservoir.
The electron distributions
in the left and the right reservoir, 
$f_1(\varepsilon)$ and $f_2(\varepsilon)$, are Fermi-Dirac
distributions. The distribution $f_{12}(\varepsilon)$ in the intermediate
reservoir depends on the type of scattering. 
}
\label{q1G.f1}
\end{figure}

The model is depicted in Fig.\ \ref{q1G.f1}. The
conductors 1 and 2 are connected via a reservoir
with distribution function $f_{12}(\varepsilon)$.
The time-averaged current $I_m$ through conductor $m=1,2$ is given by 
\begin{mathletters}
\label{q1G.1}
\begin{eqnarray}
I_1 &=& (G_1/e) \int \!\! d\varepsilon \, 
[ f_1(\varepsilon) - f_{12}(\varepsilon)]
\: , 
\label{q1G.1a} \\
I_2 &=& (G_2/e) \int \!\! d\varepsilon \, 
[f_{12}(\varepsilon) - f_2(\varepsilon)]
\: .
\label{q1G.1b}
\end{eqnarray}
\end{mathletters}%
The conductance $G_m\equiv1/R_m=G_0 \sum_{n=1}^N T_n^{(m)}$, 
with $T_n^{(m)}$ the $n$-th transmission eigenvalue of conductor $m$.
We assume small $e V$ and $k_B {\rm T}$, so that 
the energy dependence of the transmission eigenvalues can be neglected.

Current conservation requires that 
$I_1=I_2\equiv I$.
The total resistance of the conductor is given by Ohm's law,
\begin{equation}
R=R_1+R_2 \: ,
\label{q1G.4}
\end{equation}
for all three types
of scattering that we consider.
Our model is not suitable for
transport in the ballistic regime or in the quantum Hall effect regime,
where a different type of ``one-way'' reservoirs
is required \cite{but88a,but95}.

The time-averaged current (\ref{q1G.1}) depends on the
average distribution $f_{12}(\varepsilon)$ in the reservoir
between conductors 1 and 2.
In order to calculate the current fluctuations, 
we need to take into account that this distribution varies in time.
We denote the time-dependent distribution by
$\tilde{f}_{12}(\varepsilon,t)$.
The fluctuating current through conductor 1 or 2
causes electrostatic potential fluctuations $\delta \phi_{12}(t)$ 
in the reservoir, which enforce charge neutrality.
In Ref.\ \cite{b&b92}, the reservoir has a Fermi-Dirac
distribution
$\tilde{f}_{12}(\varepsilon,t)=
f[\varepsilon- E_F-  e V_{12} - e \delta \phi_{12}(t)]$, with
$E_F+eV_{12}$ the average electrochemical potential in the reservoir.
As a result, it is found that the shot-noise power $P$
of the entire conductor is
given by \cite{b&b92}
\begin{equation}
R^2 P = R_1^2 P_1 + R_2^2 P_2 \: .
\label{q1G.5}
\end{equation}
In other words, the voltage fluctuations add.
The noise powers of the two segments
depend solely on the 
time-averaged distributions \cite{but90,but92b},
\begin{equation}
P_m = 2 G_m \!\! \int \!\! d\varepsilon \, [ f_m (1-f_m) + f_{12} ( 1-f_{12})]
+ 2 S_m \!\! \int \!\! d\varepsilon \, (f_m -f_{12})^2 \: ,
\label{q1G.6}
\end{equation}
where $S_m\equiv G_0 \sum_{n=1}^N T_n^{(m)} (1 - T_n^{(m)})$.
The analysis of Ref.\ \cite{b&b92} is easily generalized to
arbitrary distribution $f_{12}$.
Then, we have 
$\tilde{f}_{12}(\varepsilon,t)=
f_{12}[\varepsilon - e \delta \phi_{12}(t)]$.
It follows that Eqs.\ (\ref{q1G.5}) and (\ref{q1G.6})
remain valid, but $f_{12}(\varepsilon)$ may be different.
Let us determine the shot noise for the three types of scattering.

{\em Quasi-elastic scattering.}
Here, it is not just the total current 
which must be conserved, but the current in each energy range.
This requires
\begin{equation}
f_{12}(\varepsilon) = 
\frac{ G_1 f_1(\varepsilon) + G_2 f_2(\varepsilon)}{G_1 + G_2} \: .
\label{q1G.8}
\end{equation}
We note that Eq.\ (\ref{q1G.8}) implies the validity of Eq.\ (\ref{q1G.4}).
Substitution of Eq.\ (\ref{q1G.8}) into Eqs.\ (\ref{q1G.5}) and (\ref{q1G.6})
yields at zero temperature the result \cite{jon96a}:
\begin{equation}
P= \left( R_1^4 S_1 + R_2^4 S_2 + R_1 R_2^2 + R_1^2 R_2
\right) R^{-3} \, P_{\rm Poisson} 
\: .
\label{q1G.9}
\end{equation}

{\em Electron heating.} 
We model electron-electron
scattering, where energy can be exchanged between the electrons
at constant total energy.
We assume that the exchange of energies establishes a 
Fermi-Dirac 
distribution $f_{12}(\varepsilon)$ 
at an  electrochemical potential $E_F + e V_{12}$
and an elevated temperature ${\rm T}_{12}$.
 From current conservation it follows that
\begin{equation}
V_{12} = (R_2/R) \, V \: .
\label{q1G.10}
\end{equation}
Conservation of the energy of the electron gas requires that
${\rm T}_{12}$
is such that no energy is
absorbed or emitted by the reservoir. 
This implies
\begin{equation}
{\rm T}_{12}^2 = {\rm T}^2 + \frac{V^2}{{\cal L}_0} \, 
\frac{R_1 R_2}{R^2}
\: ,
\label{q1G.13}
\end{equation}
with the Lorentz number 
${\cal L}_0 \equiv \case{1}{3} (\pi k_B/e)^2$.
At zero temperature in the left and right reservoir and for $R_1=R_2$
we have $k_B {\rm T}_{12}=(\sqrt{3}/2\pi) e V \simeq 0.28 e V$.
For the shot noise at ${\rm T}=0$,
we thus obtain using Eqs.\ (\ref{q1G.5}) and (\ref{q1G.6}) 
the result \cite{jon96a}:
\begin{eqnarray}
P &=& \left\{ R_1^3 S_1 + R_2^3 S_2 +
\case{1}{\pi} \sqrt{3 R_1 R_2} \vphantom{\frac{1^1}{2^2}}
\right[
R_1(1-R_1S_1) + R_2(1-R_2S_2) 
\nonumber \\
&& + \, 2 R_1^2 S_1 \ln\left(1+e^{-\pi\sqrt{R_1/3R_2}}\right) 
\nonumber \\
&& \left. \left. + \, 2
R_2^2 S_2 \ln\left(1+e^{-\pi\sqrt{R_2/3R_1}}\right) \right] \right\}
R^{-2} \, P_{\rm Poisson}
\: .
\label{q1G.14}
\end{eqnarray}

{\em Inelastic scattering.} 
The distribution function of the intermediate reservoir is 
the Fermi-Dirac distribution at the lattice
temperature ${\rm T}$, with an
electrochemical potential $\mu_{12}\equiv E_F + e V_{12}$,
where $V_{12}$ is given by Eq.\ (\ref{q1G.10}).
This reservoir absorbs energy, in contrast to the previous two cases.
The zero-temperature shot-noise power is given by \cite{b&b92}:
\begin{equation}
P =  \left(  R_1^3 S_1 + R_2^3 S_2 \right) R^{-2} \, P_{\rm Poisson}\: .
\label{q1G.16}
\end{equation}

This model will be applied to double-barrier junctions,
chaotic cavities, and disordered conductors in 
Secs.\ \ref{q2}--\ref{q4}.
Quite generally, we will find that quasi-elastic scattering 
has no effect on the shot noise, while electron heating leads to a small
enhancement of the shot noise.
Inelastic scattering suppresses the shot noise in most cases,
but not in the double-barrier junction.

\subsection{Statistics of Transmitted Charge}
\label{q1H}

The conductance is a measure for the average number of
electrons transmitted per unit time.
The shot noise quantifies the variance of the transmitted
charge.
Levitov and Lesovik 
have studied the full distribution function of charge
transmitted through a mesoscopic conductor \cite{lev93,lev96}.
This function $P_q(t)$ gives the probability that exactly
$q$ electrons have been transmitted during a given time interval $t$.
An alternative way to describe this distribution is
through its characteristic function $\chi(\lambda,t)$.
They are mutually related according to \cite{Crame}
\begin{equation}
\chi(\lambda,t)=\sum \limits_{q=0}^\infty P_q(t) e^{i q \lambda} \: ,
\;\;\;
P_q(t) = \frac{1}{2 \pi} \int \limits_{-\pi}^\pi d \lambda \,
e^{-i q \lambda} \chi(\lambda,t) \: .
\label{q1H.1}
\end{equation}
The average number of electrons transmitted during a time
$t$ is given by
\begin{equation}
\overline{q(t)} = \sum  \limits_{q=0}^\infty  q P_q(t) =
\lim \limits_{\lambda \rightarrow 0}
\frac{\partial}{i \partial \lambda} \chi(\lambda,t) \: .
\label{q1H.2}
\end{equation}
More generally, one can express the $k$-th moment $\mu_k(t)$ of the distribution
by 
\begin{equation}
\mu_k(t) \equiv \overline{q^k(t)} = \lim \limits_{\lambda \rightarrow 0}
\left( \frac{\partial}{i \partial \lambda} \right)^k \chi(\lambda,t) \: .
\label{q1H.3}
\end{equation}
The average current is simply $I=e \mu_1(t)/t$ and
the noise power equals
$P=2 e^2 \lim_{t \rightarrow \infty} \mbox{var} \, q(t) / t =
2 e^2 \lim_{t \rightarrow \infty} [\mu_2(t) - \mu_1^2(t)]/t$.

Levitov and Lesovik \cite{lev93} have computed
the characteristic function at zero temperature and at small voltage $V$.
The result
\begin{equation}
\chi_N(\lambda,t)=
\prod \limits_{n=1}^N[ (e^{i \lambda}-1)T_n + 1 ]^{G_0 V t/e} \: 
\label{q1H.6}
\end{equation}
is the characteristic function of
the binomial or Bernoulli distribution:
In scattering channel $n$, 
the charge transmitted in a time interval $t$
is due to $G_0 V t/e$ independent attempts 
to transmit an electron, each time with a probability $T_n$.
The fact that only one electron (within a single channel)
can be transmitted during a time $e/G_0 V$ is due to
the Pauli principle.
Only if $T_n \ll 1$ for all $n$, Eq.\ (\ref{q1H.6})
reduces to the characteristic function of a Poisson process.
Otherwise, the electrons are transmitted according to
sub-Poissonian statistics.

The distribution function of transmitted charge has been determined
for a normal-metal--superconductor point contact 
by Muzykantskii and Khmelnitskii \cite{muz94},
for a disordered conductor by Lee, Levitov, and Yakovets \cite{lee95}, 
and for a double-barrier junction by one of the authors \cite{jon96b}.

\section{Double-Barrier Junction}
\label{q2}

\subsection{Resonant Tunneling}
\label{q2A}

In 1973 Tsu and Esaki predicted the occurrence of a negative 
differential resistance due to resonant tunneling
through two tunnel barriers in series 
inside a semiconductor heterostructure \cite{tsu73}.
The experimental observation \cite{cha74} opened a large
field of research. 
The study of noise in resonant tunneling is a recent development,
sparked by the demonstration by Li {\em et al.}\ \cite{li90b} 
that the shot noise in an (Al,Ga)As double-barrier junction
may vary between one-half (for equal barrier heights) 
and the full Poisson noise (for very unequal barrier heights).
This observation, confirmed by other experiments 
\cite{bro91,cia95,liu95hc},
has inspired many theoreticians
\cite{che91,dav92,che92,she94,xio95,bo96,ian96}.
Below we will only consider the
zero-frequency, low-voltage limit, in order to treat the 
double-barrier junction on the same footing as the other systems
described in this review. We assume high tunnel barriers
with mode-independent transmission probabilities 
$\Gamma_1 , \Gamma_2 \ll 1$.

\begin{figure}
\centerline{\psfig{file=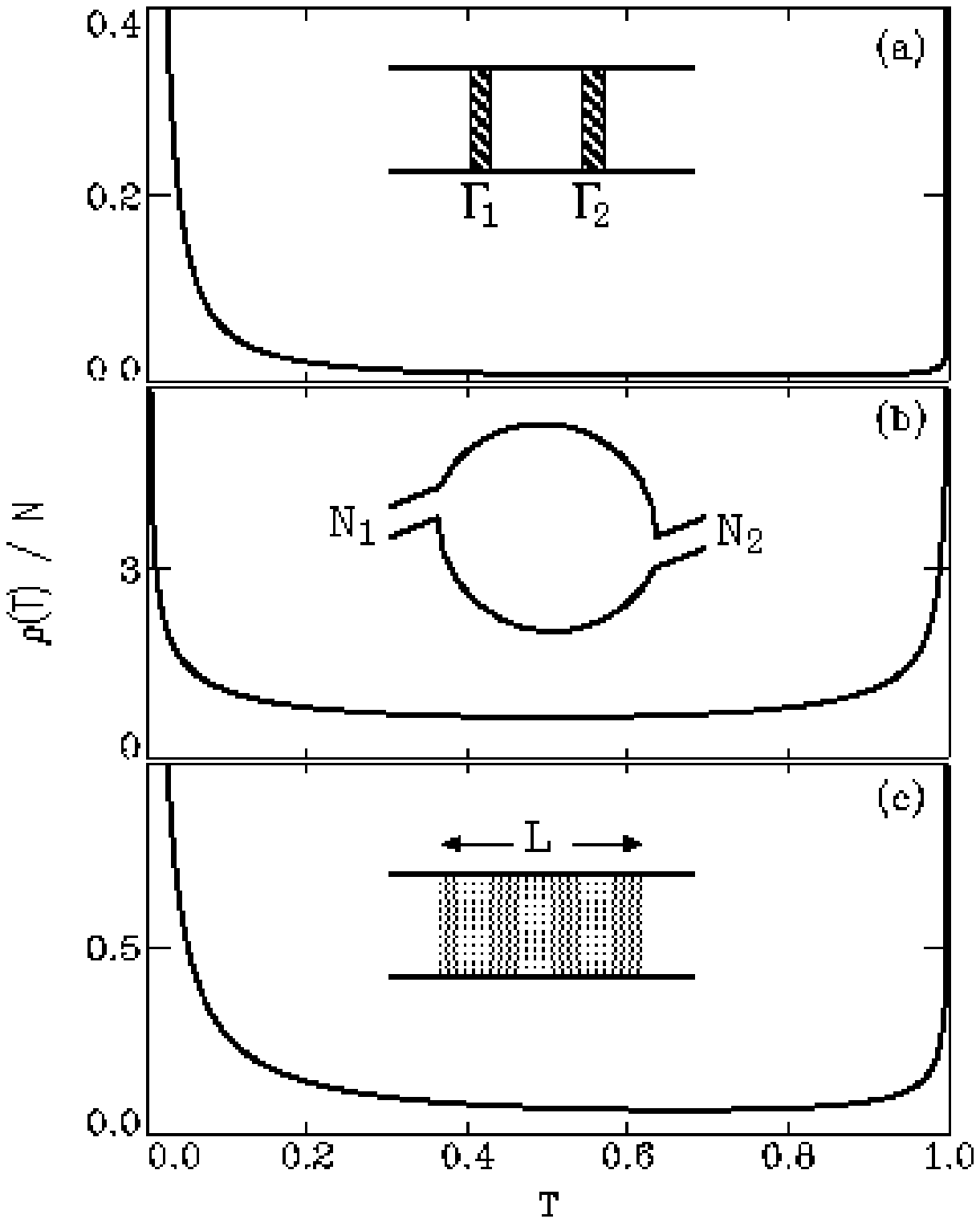,width=7cm}}
\caption{
The distribution $\rho(T)$ of transmission eigenvalues $T$
for (a) a double-barrier junction,
according to Eq.\ (\ref{q2A.2}) with $\Gamma_1=\Gamma_2=0.01$;
(b) a chaotic cavity, according to Eq.\ (\ref{rholambdaCU})
with $N_1=N_2\equiv N$;
and (c) a disordered wire,
according to Eq.\ (\ref{q4A.5}) with $L=20 \ell$.
Each structure has a bimodal distribution.
}
\label{q2A.f1}
\end{figure}

The transmission eigenvalues through the two barriers in series
are given by a Fabry-Perot type of formula,
\begin{equation}
T_n=\frac{\Gamma_1 \Gamma_2}{2 - \Gamma_1-\Gamma_2 
- 2 \sqrt{1-\Gamma_1 -\Gamma_2} \cos\phi_n }
\: ,
\label{q2A.2}
\end{equation}
where $\phi_n$ is the phase accumulated in one round trip between the
barriers. 
The density
$\rho(T)\equiv \langle \sum_{n} \delta(T-T_n) \rangle$
of the transmission eigenvalues follows from the 
uniform distribution of $\phi_n$ between 0 and $2\pi$
\cite{mel94},
\begin{equation}
\rho(T) = \frac{N \Gamma_1 \Gamma_2}{\pi (\Gamma_1+\Gamma_2)} \,
\frac{1}{\sqrt{T^3(T_{+} - T)} } \: ,
\;\;\; T \in [ T_{-}, T_{+} ] \: ,
\label{q2A.3}
\end{equation}
$\rho(T)=0$ otherwise,
with $T_{-}=\Gamma_1 \Gamma_2/ \pi^2$ and
$T_{+}=4 \Gamma_1 \Gamma_2/(\Gamma_1+\Gamma_2)^2$.
The density (\ref{q2A.3}) is plotted in Fig.\ \ref{q2A.f1}a.

The average conductance,
\begin{equation}
\langle G \rangle = G_0 \int \limits_0^1 dT \, \rho(T) \, T 
= G_0 N \frac{ \Gamma_1 \Gamma_2}{\Gamma_1 + \Gamma_2} \: ,
\label{q2A.4}
\end{equation}
is just the series conductance of the two tunnel conductances.
The resonances are averaged out by taking a uniform distribution
of the phase shifts $\phi_n$.
Physically, this averaging corresponds either to
an average over weak disorder in the region between the barriers,
or to a summation over a large number of modes if the separation
between the barriers is large compared to the Fermi wave length,
or to an applied voltage larger than the width of the resonance.

For the shot-noise power one obtains
\begin{equation}
\langle P \rangle
= P_0 \int \limits_0^1 dT \, \rho(T) \, T(1-T)=
\frac{ \Gamma_1^2 + \Gamma_2^2}{(\Gamma_1 + \Gamma_2)^2}
P_{\rm Poisson} \: ,
\label{q2A.5}
\end{equation}
using Eqs.\ (\ref{q2A.3}) and (\ref{q2A.4}).
This result was first derived by Chen and Ting \cite{che91}.
For asymmetric junctions, one barrier dominates the
transport and the shot noise equals the Poisson noise.
For symmetric junctions, the shot noise gets suppressed
down to $\langle P \rangle = \case{1}{2} P_{\rm Poisson}$ for
$\Gamma_1=\Gamma_2$.
The theoretical result (\ref{q2A.5}) is in agreement with the experimental
observations \cite{li90b,bro91,cia95,liu95hc}.

The suppression of the shot noise below $P_{\rm Poisson}$ 
in symmetric junctions is a consequence
of the {\em bimodal} distribution of transmission eigenvalues,
as plotted in Fig.\ \ref{q2A.f1}a. 
Instead of all $T_n$'s being close to the average transmission
probability, the $T_n$'s are either close
to 0 or to 1. This reduces the sum $T_n(1-T_n)$.
A similar suppression mechanism exists for shot noise
in chaotic cavities and in disordered conductors, see
Secs.\ \ref{q3} and \ref{q4}.

Phase coherence is not essential
for the occurrence of suppressed shot noise.
Davies {\em et al.}\ obtained the result (\ref{q2A.5})
from a model of incoherent sequential tunneling \cite{dav92}.
The method of Sec.\ \ref{q1G}
(with $G_m=S_m=G_0 N \Gamma_m$ for $m=1,2$) shows that
both quasi-elastic scattering [see Eq.\ (\ref{q1G.9})]
and inelastic scattering [see Eq.\ (\ref{q1G.16})] do not modify
Eq.\ (\ref{q2A.5}).
Thermalization of the electrons in the region between the barriers
enhances the shot noise, as follows
from Eq.\ (\ref{q1G.14}).
For $\Gamma_1=\Gamma_2$
we find
\begin{equation}
P = \left[ \case{1}{2} + \case{\sqrt{3}}{\pi}
\ln \left( 1 + e^{-\pi/\sqrt{3}} \right) \right] \, P_{\rm Poisson} 
\simeq 0.58 P_{\rm Poisson} \: ,
\label{q2A.6}
\end{equation}
which is slightly above the one-half suppression in the
absence of thermalization.
More theoretical work on the influence of internal scattering and
of dephasing on the shot noise in
double-barrier junctions is contained in Refs.\
\cite{ala92,car94,gal94,dav95,naz96}.

\subsection{Coulomb Blockade}
\label{q2B}

The suppression of the shot noise described in the previous Section
is due to correlations induced by the Pauli principle.
Coulomb interactions are another
source of correlations among the electrons.
A measure of the importance of Coulomb repulsion is the
charging energy $E_C=e^2/2C$ of a single electron inside
the conductor with a capacitance $C$.
In open conductors, where $C$ is large, charging effects
are expected to be negligible \cite{but93}.
For closed conductors, such as a double-barrier junction, $E_C$ 
can be larger than $k_B$T, in which case charging effects
have a pronounced influence on the conduction \cite{aveMP}.
If $e V < E_C$,
conduction through the junction is suppressed. This is known as the
Coulomb blockade.
At $eV > E_C$,
one electron at a time
can tunnel into the junction. 
The next electron can follow, only after the
first electron has tunneled out of the junction. This is the
single-electron tunneling regime.
The theory of shot noise in single-electron tunneling 
devices has been developed
by Korotkov {\em et al}.\ \cite{kor92}, 
Hershfield {\em et al}.\ \cite{her93}, 
and others \cite{levy93,hun93,han93,ima93,kre93,kor94,mul96}.

\begin{figure}
\centerline{\psfig{file=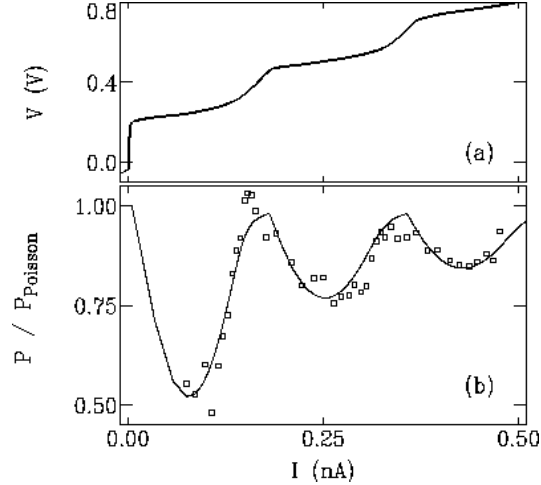,width=7cm}}
\caption{
Experimental results by Birk {\em et al}.\ \protect\cite{bir95}
in the single-electron tunneling regime.
The double-barrier junction consists of a 
tip positioned above a nanoparticle on a substrate.
(a) Experimental voltage $V$ versus current $I$.
(b) Shot-noise power $P$ versus $I$. Squares: experiment; solid line: 
theory of Hershfield {\em et al}.\ \protect\cite{her93}.}
\label{q2B.f1}
\end{figure}

Experiments have been reported
by Birk, de Jong, and Sch\"{o}nenberger \cite{bir95}.
Here,
the double-barrier junction was formed by a scanning-tunneling
microscope positioned above a metal nanoparticle on an
oxidized substrate. Due to the small size of the particle,
$E_c \geq 1000 k_B$T, at ${\rm T} = 4\:$K\@.
The relative heights of the two tunnel barriers
can be modified by changing the tip-particle distance.
Experimental results for an asymmetric junction
are plotted in Fig.\ \ref{q2B.f1}.
The $I$-$V$ characteristics display a stepwise increase of the
current with the voltage.
(Rotating the plot 90$^\circ$ yields the usual
presentation of the `Coulomb staircase.')
At small voltage, $I\simeq 0$ due to to the Coulomb blockade.
At each subsequent step in $I$, the number of excess electrons in the 
junction increases by one.
The measured shot noise oscillates 
along with the step structure in the $I$-$V$ curve.
The full shot-noise level $P=P_{\rm Poisson}$ is reached at each 
plateau of constant $I$.
In between, $P$ is suppressed down to $\case{1}{2} P_{\rm Poisson}$.
The experimental data are in excellent agreement with the theory of
Ref.\ \cite{her93}.

A qualitative understanding of the periodic shot-noise suppression
caused by the Coulomb blockade goes as follows:
On a current plateau in the $I$-$V$ curve, the number of electrons
in the junction is constant for most of the time. 
Only during a very short instance
an excess electron occupies the junction, leading to the transfer of
one electron. This fast transfer process is dominated by the highest
tunnel barrier. Since the junction is asymmetric, Poisson noise is
expected.
The situation is different for voltages where there is
a step in the $I$-$V$ curve. Here, two charge states are 
degenerate in total energy. 
If an electron tunnels into the junction, it may stay 
for a longer time, during which tunneling of the next
electron is forbidden. Both barriers are thus alternately blocked.
This leads to a correlated current,
yielding a suppression of the shot noise.

An essential requirement for the Coulomb blockade is that 
$G \lesssim e^2/h$. 
For larger $G$  the quantum-mechanical charge fluctuations
in the junction become big enough to overcome the Coulomb blockade.
The next Section will discuss shot noise in a quantum dot,
without including Coulomb interactions. This is justified as long as
$G \gtrsim e^2/h$. For smaller $G$, the quantum dot behaves essentially as the
double-barrier junction considered above.

\section{Chaotic Cavity}
\label{q3}

A cavity of sub-micron dimensions, etched in a semiconductor is called a
quantum dot. The transport properties of the quantum dot can be measured by
coupling it to two electron reservoirs, and bringing them out of equilibrium.
We consider the generic case that the classical motion in the cavity can be
regarded as chaotic, as a result of scattering by randomly placed impurities or
by irregularly shaped boundaries. Then transport quantities are insensitive to
microscopic properties of the quantum dot, such as the shape of the cavity and
the degree of disorder.

A theory of transport through a chaotic cavity can be based on the single
assumption that the scattering matrix of the system is uniformly distributed in
the unitary group \cite{Bar94,Jal94}. This is the ``circular ensemble'' of
random-matrix theory \cite{Meh91,Bee97}. The assumption of a uniform distribution of
the scattering matrix is valid if the coupling to the electron reservoirs
occurs via two {\em ballistic\/} point contacts, with a conductance
$G_m=G_{0} N_m$, $m=1,2$. (Otherwise a more general distribution, known as the
``Poisson kernel'' applies \cite{Bro95}.) The presence of time-reversal
symmetry is accounted for by restricting the scattering matrix to the subset of
symmetric unitary matrices. This is known as the circular orthogonal ensemble
(labeled by the index $\beta=1$). If any unitary matrix is equally probable,
the ensemble is called circular unitary ($\beta=2$).

To compute the statistics of transport properties in a quantum dot one needs to
know the distribution of the transmission eigenvalues in the circular ensemble.
In the most general case $N_{1} \neq N_{2}$, 
the transmission matrices $t_{12}$ and $t_{21}$
are rectangular. The two matrix products
$t_{12}^{\vphantom{\dagger}}t_{12}^{\dagger}$ and
$t_{21}^{\vphantom{\dagger}}t_{21}^{\dagger}$ contain a common set of ${\rm
min}(N_{1},N_{2})$ non-zero transmission eigenvalues. Only these contribute to
the transport properties. 
For $N_{i}\gg 1$ the distribution $\rho(T)$ of the 
transmission eigenvalues is \cite{Naz95}
\begin{equation}
\rho(T)=\frac{1}{\pi}(N_{1}N_{2})^{1/2}\frac{1}{T}
\left[\frac{T-T_{-}}{(1-T)(1-T_{-})}\right]^{1/2} \: ,
\;\;\; T \in [T_{-},1] \: ,
\label{rholambdaCU}
\end{equation}
$\rho(T)=0$ otherwise, with 
$T_{-}=(N_1-N_2)^2/(N_1^2 + N_2^2)$.
This density is plotted in Fig.\ \ref{q2A.f1}b.

The average conductance,
\begin{equation}
\langle G \rangle
=G_0 \int_0^1
\!dT\,\rho(T)\,T=
G_0 \frac{N_{1}N_{2}}{N_{1}+N_{2}} \: ,
\label{Gaverage}
\end{equation}
is the series conductance of the two point contacts.
The average shot-noise power,
\begin{equation}
\langle P\rangle
=P_0\int_0^1
\!dT\,\rho(T)\,T(1-T)= \frac{N_{1}N_{2}}{(N_{1}+N_{2})^{2}} \, P_{\rm Poisson} 
\: ,
\label{PoverG}
\end{equation}
is smaller than the Poisson noise.
For two identical point contacts the suppression factor is one quarter
\cite{Jal94}, to be compared with the one-half suppression
in a double-barrier junction (see Sec.\ \ref{q2A}) and the
one-third suppression in a
disordered wire (see Sec.\ \ref{q4A}).

The result (\ref{PoverG}) does not require
phase coherence, as follows from Eq.\ (\ref{q1G.9})
using $S_m=0$ for $m=1,2$.
However, it is affected by thermalization of the electrons and also
by inelastic scattering.
From Eq.\ (\ref{q1G.14}) we find in the case of complete
thermalization,
\begin{equation}
P = \frac{\sqrt{ 3 N_1 N_2 }}{\pi ( N_1 + N_2 )} \, P_{\rm Poisson} \: .
\label{PoverGthermo}
\end{equation}
For $N_{1}=N_{2}$, this yields 
$P = (\sqrt{3}/2\pi) P_{\rm Poisson} \simeq 0.28 P_{\rm Poisson}$.
As follows from Eq.\ (\ref{q1G.16}), inelastic scattering 
suppresses the shot noise completely.

\section{Disordered Metal}
\label{q4}
\subsection{One-third Suppression}
\label{q4A}

We now turn to transport through a diffusive conductor of length
$L$ much greater than the mean free path
$\ell$, in the metallic regime 
($L \ll$ localization length).
The average conductance is given by
the Drude formula,
\begin{equation}
\langle G \rangle
=G_0 \, \frac{N \ell}{L} \: ,
\label{q4A.3}
\end{equation}
up to small corrections of order $G_0$ (due to weak localization).
The mean free path $\ell=a_d \, \ell_{\rm tr}$
equals the transport mean free path $\ell_{\rm tr}$ times 
a numerical coefficient, which
depends on the dimensionality $d$ of the Fermi surface
($a_2 = \pi/2$, $a_3=4/3$).

 From Eq.\ (\ref{q4A.3}) one might surmise that
for a diffusive conductor all the transmission eigenvalues are
of order $\ell /L$, and hence $\ll 1$.
This would imply the shot-noise power $P=P_{\rm Poisson}$ of
a Poisson process. 
This surmise is completely
incorrect, as was first pointed out by Dorokhov \cite{dor84}, 
and later by Imry \cite{imr86} and by Pendry, MacKinnon, 
and Roberts \cite{pen92}.
A fraction
$\ell/L$ of the transmission eigenvalues is of order unity
(open channels), 
the others being exponentially small (closed channels).
For $\ell \ll L \ll N \ell$,
the density of the $T_n$'s is given by \cite{dor84}
\begin{equation}
\rho(T) = 
\frac{N \ell}{2 L} \, 
\frac{1}{T \sqrt{1-T}} \: ,
\;\;\; T \in [T_{-},1] \: ,
\label{q4A.5}
\end{equation}
$\rho(T)=0$ otherwise,
with $T_{-}=4 e^{-2 L / \ell}$.
The density $\rho(T)$, plotted in Fig.\ \ref{q2A.f1}c,
is again bimodal with peaks near unit and zero transmission.
Dorokhov \cite{dor84} obtained Eq.\ (\ref{q4A.5}) 
from a scaling
equation, which describes the evolution
of $\rho(T)$ on increasing $L$ in a wire geometry \cite{dor82,mel88a}.
A derivation for other geometries has been given
by Nazarov \cite{naz94}.

One easily checks that the bimodal distribution (\ref{q4A.5}) 
leads to the Drude conductance (\ref{q4A.3}).
For the average shot-noise power it implies
\begin{equation}
\langle P \rangle =
P_0 \, \frac{N \ell}{3 L} = \frac{1}{3} P_{\rm Poisson}
\; .
\label{q4A.6}
\end{equation}
This suppression of the shot noise
by a factor one-third is {\em universal\/},
in the sense that it does not depend on the specific geometry nor
on any intrinsic material parameter (such as $\ell$).
The one-third suppression was discovered by 
Beenakker and B\"{u}ttiker \cite{b&b92}
in the way described above, and by others
using different methods \cite{nag92,naz94,jon92,alt94}.
Nagaev's theory \cite{nag92}
is based on the semiclassical Boltzmann-Langevin equation, 
see Sec.\ \ref{q1F}.
One might therefore infer that there is
also a semiclassical derivation of  
the bimodal distribution of transmission
eigenvalues, which is the key ingredient of the quantum-mechanical
theory. 
Such a derivation is given in Ref.\ \cite{mor}.

\subsection{Dependence on wire length}
\label{q4B}

\begin{figure}
\centerline{\psfig{file=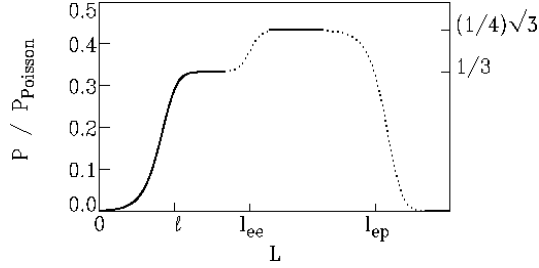,width=7cm}}
\caption{
The shot-noise power $P$ of a disordered metallic wire as a function of its
length $L$, as predicted by theory. 
Indicated are the elastic mean free path $\ell$,
the electron-electron scattering length $l_{ee}$ and the electron-phonon
scattering length $l_{ep}$.
Dotted lines are interpolations
(after Ref.\ \protect\cite{ste96}).
}
\label{q4B.f2}
\end{figure}

The one-third suppression of the shot noise breaks down if the
conductor becomes too short or too long.
Upon decreasing the length of the conductor, when $L$ becomes comparable to
$\ell$, the electron transport is no longer diffusive, but enters
the ballistic regime. 
Then the shot noise is suppressed more strongly,
according to \cite{jon92}
\begin{equation}
P = \case{1}{3} 
\left[ 1 -
(1 + L/\ell)^{-3} \right]  P_{{\rm Poisson}}
\: .
\label{q4Ba.1}
\end{equation} 
For $L \ll \ell$ there is no shot noise, as in a ballistic point contact
\cite{kul84}.
Equation (\ref{q4Ba.1}) is exact for a special model of one-dimensional
scattering, but holds more generally within a few
percent \cite{jon96a}.
A Monte-Carlo simulation in a wire geometry \cite{liu95b}
is in good agreement with Eq.\ (\ref{q4Ba.1}).
The crossover of the shot noise from the ballistic to the
diffusive regime is plotted in Fig.\ \ref{q4B.f2}.
Upon increasing $L$ at constant cross section of the conductor, 
one enters the localized regime.
Here, even the largest transmission eigenvalue 
is exponentially small \cite{dor84}, so that
$P=P_{\rm Poisson}$.
Shot noise in one-dimensional chains
for various models of disorder has been
studied in Ref.\ \cite{mel95}.

Experimentally, the crossover from the metallic to the localized regime
is usually not reached,
because phase coherence is broken when $L$ is still much smaller
than the localization length $N \ell$.
In the remainder of this Section, we apply the method of Sec.\ \ref{q1G}
to determine the effect of phase breaking and other
inelastic scattering events on the shot noise
in a disordered metal \cite{jon96a}.
We divide the conductor
into $M$ segments connected by reservoirs,
taking the 
continuum limit $M \rightarrow \infty$.
The electron distribution at position $x$ is denoted by
$f(\varepsilon,x)$.
At the ends of the conductor $f(\varepsilon,0)=f_1(\varepsilon)$ and
$f(\varepsilon,L)=f_2(\varepsilon)$, {\em i.e.}\
the electrons are
Fermi-Dirac distributed at temperature ${\rm T}$
and with electrochemical potential $\mu(0) = E_F + e V$ and
$\mu(L)=E_F$, respectively.
It follows from Eqs.\ (\ref{q1G.5}) and (\ref{q1G.6}) that
the noise power is given by 
\begin{equation}
P= \frac{4}{R L} \int \limits_0^L \! d x \! 
\int  \limits_0^\infty \! d \varepsilon 
\, f(\varepsilon,x) [1 - f(\varepsilon,x)] \: ,
\label{q4B.4}
\end{equation}
a formula first obtained by Nagaev \cite{nag92}.
We evaluate Eq.\ (\ref{q4B.4}) for the three types of scattering
discussed in Sec.\ \ref{q1G}.

\begin{figure}[tb]
\centerline{\psfig{file=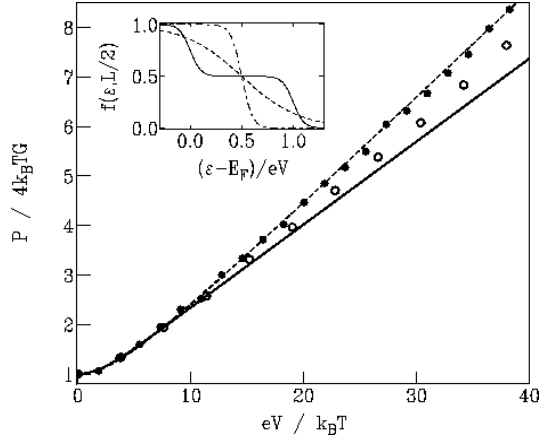,width=7cm}}
\caption{
The noise power $P$  versus voltage $V$ for a
disordered wire in the presence of quasi-elastic scattering 
[solid curve, from Eq.\ (\ref{q4B.7})] and
of electron heating [dashed curve, from Eq.\ (\ref{q4B.17})].
The inset gives the electron distribution
in the middle of the wire at $k_B {\rm T}= \case{1}{20} e V$.
The distribution for inelastic scattering is included for
comparison (dash-dotted).
Experimental data of Steinbach, Martinis, and Devoret 
\protect\cite{ste96} on silver wires at ${\rm T} = 50\:$mK
are indicated for length $L=1 \mu$m (circles) and $L=30 \mu$m (dots).
}
\label{q4B.f1}
\end{figure}

{\em Quasi-elastic scattering.} 
Current conservation and the absence of inelastic scattering
requires
\begin{equation}
f(\varepsilon,x) = 
\frac{L-x}{L} f(\varepsilon,0) + \frac{x}{L} f(\varepsilon,L)
\: .
\label{q4B.6}
\end{equation}
The electron distribution at $x=L/2$ is plotted in the inset of
Fig.\ \ref{q4B.f1}.
Substitution of Eq.\ (\ref{q4B.6})
into Eq.\ (\ref{q4B.4}) yields \cite{nag92}
\begin{equation}
P=\case{2}{3} \left[ 4 k_B {\rm T} G  + e I \coth(eV/2 k_B {\rm T}) \right] \: .
\label{q4B.7}
\end{equation}
At zero temperature the shot noise is one-third of the Poisson noise.
The same result follows from the phase-coherent theory
[Eqs.\ (\ref{q1D.21}) and (\ref{q4A.5})],
demonstrating that quasi-elastic scattering has no effect on the
shot noise. 
The temperature dependence of $P$ is plotted in Fig.\ \ref{q4B.f1}.

{\em Electron heating.}
The electron-distribution function 
is a Fermi-Dirac distribution with a spatially dependent
electrochemical potential $\mu(x)$ and temperature T$_e(x)$,
\begin{mathletters}
\label{q4B.b}
\begin{eqnarray}
f(\varepsilon,x) &=&
\left\{ 1+ \exp \left[ \frac{\varepsilon-\mu(x)}{k_B {\rm T}_e(x)} \right]
\right\}^{-1} \: ,
\label{q4B.8} \\
\mu(x) &=& E_F + \frac{L-x}{L} e V \: ,
\label{q4B.11} \\
{\rm T}_e(x)& =&\sqrt{ {\rm T}^2 + (x/L)[1-(x/L)] \, V^2 /{\cal L}_0 }
\: ,
\label{q4B.15}
\end{eqnarray}
\end{mathletters}%
cf.\ Eqs.\ (\ref{q1G.10}) and (\ref{q1G.13}).
Equations (\ref{q4B.4}) and (\ref{q4B.b})
yield
for the noise power the result \cite{nag95,koz95,dev}
\begin{equation}
P= 2 k_B {\rm T} G + 2 e I \left[ \frac{2\pi}{\sqrt{3}} 
\left( \frac{k_B {\rm T}}{e V} \right)^2 + \frac{\sqrt{3}}{2\pi} \right]
\arctan \left( \frac{\sqrt{3}}{2\pi} \, \frac{e V}{k_B {\rm T}} \right)
\: ,
\label{q4B.17}
\end{equation}
plotted in Fig.\ \ref{q4B.f1}.
In the limit $eV \gg k_B {\rm T}$ one finds
\begin{equation}
P= \case{1}{4} \sqrt{ 3 } \, P_{\rm Poisson} \simeq  0.43 \, P_{\rm Poisson}
\: .
\label{q4B.18}
\end{equation}
Electron-electron scattering increases the shot noise above
$\case{1}{3} P_{\rm Poisson}$ because the
exchange of energies makes the current less
correlated.

{\em Inelastic scattering.} 
The electron-distribution function is given by
\begin{equation}
f(\varepsilon,x)=
\left\{ 1+ \exp \left[ \frac{\varepsilon-\mu(x)}{k_B {\rm T}} \right]
\right\}^{-1} \: ,
\label{q4B.19}
\end{equation}
with $\mu(x)$ according to Eq.\ (\ref{q4B.11}).
We obtain from Eqs.\ (\ref{q4B.4}) and (\ref{q4B.19})
that the noise power
is equal to the Johnson-Nyquist noise (\ref{q1B.2})
for arbitrary $V$. 
The shot noise is thus
completely suppressed by inelastic scattering
\cite{nag95,koz95,shi92,b&b92,lan93,liu94}.

The dependence of the shot-noise power on the length of a 
disordered conductor is plotted in Fig.\ \ref{q4B.f2}.
The phase coherence length (between $\ell$ and $l_{ee}$)
does not play a role.
An early experimental demonstration of sub-Poissonian shot noise
in a wire defined in an (Al,Ga)As heterostructure
was reported by Liefrink {\em et al.} \cite{lie94}.
The measurements
were in agreement with theory, but lacked the precision needed to discriminate
between the elastic and the hot-electron value.
More accurate experiments by Steinbach, Martinis, and Devoret \cite{ste96}
on silver wires are shown in Fig.\ \ref{q4B.f1}. 
The noise in a wire of $L=30 \: \mu$m is in excellent agreement
with the hot-electron result (\ref{q4B.17}). For the $L=1\:\mu$m wire
the noise crosses over to the elastic result (\ref{q4B.7}),
without quite reaching it.

\section{Aharonov-Bohm Effect}
\label{q5}

Since the current operator is a one-particle observable, the shot-noise power
(given by the current-current correlator) is a {\em two-particle\/} transport
property. This is an essential difference with the conductance, which is a
one-particle property. The distinction between one-particle and two-particle
properties is relevant even without Coulomb interactions between the electrons,
because of the quantum-mechanical exchange interaction. A striking
demonstration of the two-particle nature of shot noise, discovered by
B\"{u}ttiker \cite{but91,but92}, occurs in the Aharonov-Bohm effect.

\begin{figure}
\centerline{
\psfig{figure=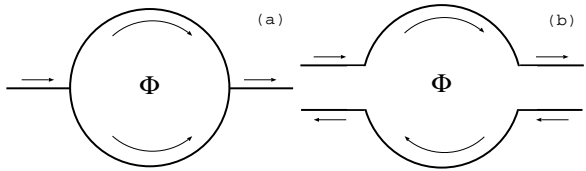,width=8cm}
}%
\medskip
\caption{
Feynman paths enclosing a magnetic flux $\Phi$. The paths in (a) correspond to
the same electron, the paths in (b) to two different electrons.
}\label{ABfig}
\end{figure}

The Aharonov-Bohm effect in electrical conduction is a periodic oscillation of
the conductance of a ring (or cylinder) as a function of the enclosed magnetic
flux $\Phi$. (For reviews, see Refs.\ \cite{was86,aro87}.) The
fundamental periodicity of the oscillation is $h/e$, because a flux increment
of an integer number of flux quanta changes by an integer multiple of $2\pi$
the phase difference between Feynman paths along the two arms of the ring (see
Fig.\ \ref{ABfig}a). Since the conductance is a one-particle property, the two
interfering Feynman paths must belong to the same electron, which on entering
the ring has a probability to traverse the ring either clock-wise or
counter-clock-wise. For a maximal amplitude of the conductance oscillations the
two probabilities should be approximately equal. The Lorentz force causes an
electron to traverse the ring preferentially in one of the two directions.
This is why the Aharonov-Bohm effect is suppressed by a strong magnetic field
\cite{C&H}.

Shot noise can exhibit Aharonov-Bohm oscillations which persist in a strong
magnetic field. The Lorentz force guides the right-moving electron along the
upper arc of the ring, and the left-moving electron along the lower arc (see
Fig.\ \ref{ABfig}b). The absolute value squared $|\psi_{1}|^{2}$,
$|\psi_{2}|^{2}$ of each of the two single-particle wave functions
$\psi_{1},\psi_{2}$ does not depend on $\Phi$, hence the conductance is
$\Phi$-independent. However, the absolute value squared of the two-particle
wave function $|\Psi({\bf r}_{1},{\bf r}_{2})|^{2}=|\psi_{1}({\bf
r}_{1})\psi_{2}({\bf r}_{2})-\psi_{1}({\bf r}_{2})\psi_{2}({\bf r}_{1})|^{2}$
does depend on $\Phi$. The flux sensitivity is an exchange effect, which
vanishes only if the two single-particle wave functions do not overlap.

To measure the flux sensitivity in the shot noise 
due to the exchange effect one can not simply
use the ring geometry of Fig.\ \ref{ABfig}: Because each electron is
fully transmitted the shot noise vanishes completely
in a strong magnetic field.
(The case of weak magnetic field has been studied in Ref.\ \cite{dav94}.)
B\"uttiker
\cite{but92} has suggested a four-terminal configuration, where the
correlator of the current at two terminals is measured using the other
two terminals as current sources. Lesovik and Levitov \cite{les94} have
proposed a two-terminal configuration, but with a time-dependent
magnetic field. Ideally, the shot noise should show a flux sensitivity
while the time-averaged current should not. Observation of this effect
remains an experimental challenge.

\section{Cooper Pairs}
\label{q6}

\subsection{Normal-Metal--Superconductor Junctions}
\label{q6A}

If a normal metal is connected to a superconductor,
the dissipative normal current is converted into dissipationless
supercurrent. This conversion goes through a process 
called Andreev reflection \cite{and64}: Incoming electrons are reflected 
into outgoing holes, with the transfer of a Cooper pair into the
superconductor.
Since the elementary charge transfer now involves a charge $2e$
instead of $e$, one might expect a doubling of the shot noise in an
NS junction. Let us see how this follows from the theory 
\cite{khl87,muz94,jon94b}.

We assume low temperatures and an applied voltage $e V$
smaller than the  excitation gap $\Delta$ in the superconductor,
so that the electrons and holes are confined to the normal metal. 
The scattering from incoming 
into outgoing states 
is described by the
$2 N \times 2 N$ reflection matrix $\sf R$,
\begin{equation}
\left( \begin{array}{c} O_e \\ O_h \end{array} \right) =
{\sf R} 
\left( \begin{array}{c} I_e \\ I_h \end{array} \right)
\: , \;\;\;
{\sf R} = 
\left( \begin{array}{ll}
{\sf r}_{\vphantom{h}ee} & {\sf r}_{eh} \\
{\sf r}_{he} & {\sf r}_{hh} 
\end{array} \right)
\: ,
\label{q6A.01}
\end{equation}
where $I_e,I_h,O_e,O_h$ are the $N$-component vectors denoting the
amplitudes of
the incoming ($I$) and outgoing ($O$) electron ($e$) 
and hole ($h$) modes.
The reflection matrix ${\sf R}$ can be decomposed in
$N \times N$ submatrices, where {\em e.g.}\
${\sf r}_{he}$ contains the reflection amplitudes from incoming
electrons into outgoing holes.
The conductance \cite{btk82,lam91,tak92}
and the shot-noise power \cite{jon94b} are given by
\begin{eqnarray}
G_{{\rm NS}} &=& 2 G_0  \mbox{Tr} \, 
{\sf r}^{\vphantom{\dagger}}_{he} {\sf r}_{he}^\dagger 
= 2 G_0 \sum \limits_{n=1}^N {\cal R}_n
\: ,
\label{q6A.1}
\\
P_{{\rm NS}} &=& 4 P_0  \mbox{Tr} 
\, {\sf r}_{he}^{\vphantom{\dagger}} {\sf r}_{he}^\dagger 
({\sf 1} -  {\sf r}_{he}^{\vphantom{\dagger}} {\sf r}_{he}^\dagger )
= 4 P_0 \sum \limits_{n=1}^N {\cal R}_n ( 1 - {\cal R}_n )
 \: ,
\label{q6A.2}
\end{eqnarray}
with ${\cal R}_n$ an eigenvalue of 
${\sf r}_{he}^{\vphantom{\dagger}} {\sf r}_{he}^\dagger$,
evaluated at the Fermi energy.

The eigenvalue ${\cal R}_n$ can be related
to the scattering properties of the
normal region through the
Bogo\-liubov-de Gennes equation \cite{SM&A},
which is a $2\times 2$ matrix Schr\"{o}dinger equation
for electron and hole wave functions.
In the presence of time-reversal symmetry, 
${\cal R}_n$ can be expressed in terms of
the transmission eigenvalue $T_n$ of the normal region \cite{bee92}:
\begin{equation}
{\cal R}_n = T_n^2 (2 - T_n)^{-2} \: .
\label{q6A.3}
\end{equation}
Substitution into Eqs.\ (\ref{q6A.1}) and (\ref{q6A.2})
yields \cite{jon94b,bee92}
\begin{eqnarray}
G_{{\rm NS}}&=& G_0 \sum \limits_{n=1}^N \frac{2 T_n^2}{(2 - T_n)^2 } \: ,
\label{q6A.4} \\
P_{{\rm NS}} &=& P_0 \sum \limits_{n=1}^N 
\frac{16 T_n^2 (1 - T_n)}{(2 - T_n)^4} \: .
\label{q6A.5}
\end{eqnarray}
As in the normal state, scattering channels which have 
$T_n=0$ or $T_n=1$ do not contribute to the shot noise.
However, the way in which partially transmitting channels contribute
is entirely different from the normal state result (\ref{q1D.20}).

\begin{figure}[tb]
\centerline{\psfig{file=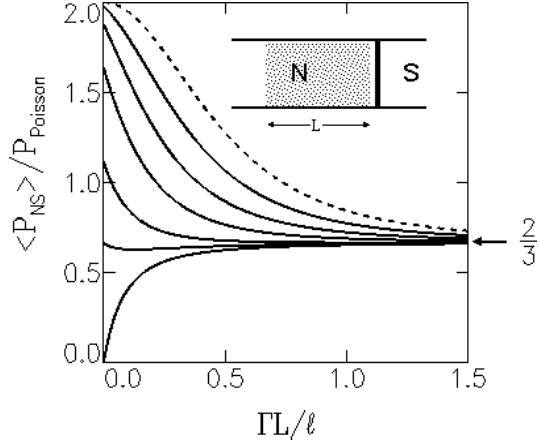,width=7cm}}
\caption{The shot-noise power $\langle P_{\rm NS} \rangle$
of a disordered NS junction 
with a barrier at the NS interface (shown in the inset)
as a function of
its length $L$, for barrier transparencies 
$\Gamma= 1,0.9,0.8,0.6,0.4,0.2, \ll 1$
from bottom to top. 
For $L=0$, $\langle P_{{\rm NS}} \rangle$ varies with $\Gamma$ 
according to Eq.\ (\protect\ref{q6A.6}).
If $L$ increases it approaches the limiting value
$\langle P_{{\rm NS}} \rangle=\frac{4}{3} e I$ for 
each $\Gamma$
(after Ref.\ \protect\cite{jon94b}).}
\label{q6A.f2}
\end{figure}

For a planar tunnel barrier ($T_n=\Gamma$ for all $n$)
one finds \cite{khl87}
\begin{equation}
P_{{\rm NS}} = P_0 N \frac{16 \Gamma^2 (1 - \Gamma)}{(2 - \Gamma)^4} =
\frac{8 (1 - \Gamma)}{(2 - \Gamma)^2} P_{{\rm Poisson}} 
\: ,
\label{q6A.6}
\end{equation}
which for $\Gamma \ll 1$ simplifies to
$P_{{\rm NS}}= 4 e I = 2 P_{{\rm Poisson}}$. 
This can be interpreted as an uncorrelated
current of $2e$-charged particles.

Since Eq.\ (\ref{q6A.5}) is valid for arbitrary scattering region,
we can easily determine the average shot-noise power
for a double-barrier junction in series
with a superconductor,
\begin{equation}
\langle P_{\rm NS} \rangle= 
\left( 2 - \frac{5 \Gamma_1^2 \Gamma_2^2}{(\Gamma_1^2 + \Gamma_2^2)^2}
\right) P_{\rm Poisson} \: ,
\label{q6A.8}
\end{equation}
for a chaotic cavity in series with a superconductor,
\begin{equation}
\langle P_{\rm NS}  \rangle= 
\frac{x^2}{ (1+x)^2 ( \sqrt{1+x} -1 )}
P_{\rm Poisson} \: ,
\label{q6A.9}
\end{equation}
with $x=4 N_1 N_2/(N_1+N_2)^2$,
and for a disordered NS junction \cite{jon94b},
\begin{equation}
\langle P_{\rm NS} \rangle= 
\frac{2}{3}
P_{\rm Poisson} \: .
\label{q6A.10}
\end{equation}
The average was computed with the densities 
(\ref{q2A.3}), (\ref{rholambdaCU}),
and (\ref{q4A.5}) of transmission eigenvalues, respectively.
The shot noise in a disordered NS junction
with a barrier at the NS interface is plotted in Fig.\
\ref{q6A.f2}. It makes the connection between the results 
(\ref{q6A.6}) and (\ref{q6A.10}).
We have indeed found that for a high tunnel barrier and for a disordered
NS junction the shot noise is doubled with respect to the
normal-state results. For other systems, the relation is
more complicated.

More theoretical work on shot noise in NS systems is given in
Refs.\ \cite{hes96,ana96,mar96b}.
The effects of the Coulomb blockade
on the shot noise in low-capacitance NSN junctions
are described in Refs.\ \cite{han94,kre94}.
With one inconclusive  exception \cite{vys83},
no experimental observation of shot noise in NS junctions has been
reported, yet.

\subsection{Josephson Junctions}
\label{q6B}

A Josephson junction contains two normal-metal--superconductor interfaces, with
a phase difference $\phi$ of the superconducting order parameter. Such an SNS
junction sustains a current $I(\phi)$ in equilibrium, {\em i.e.}\ even if the voltage
difference $V$ between the superconductors vanishes. Since this supercurrent is
a ground-state property, it can not fluctuate by itself. (It exhibits no shot
noise.) Time-dependent fluctuations result from quasiparticles which are
excited at any finite temperature $T$. Their zero-frequency power density
$P(\phi)$ is related to the linear-response conductance
$G(\phi)=\lim_{V\rightarrow 0}\partial I(\phi)/\partial V$ of the Josephson
junction in the same way as in the normal state,
\begin{equation}
P(\phi)=4k_{\rm B}TG(\phi) \: ,
\label{PGphi}
\end{equation}
cf.\ Eq.\ (\ref{q1B.2}). 
A remarkable difference with Johnson-Nyquist noise in a
normal metal is that $P(\phi)$ may actually increase with decreasing
temperature, because of the rapid increase of $G(\phi)$ when $T\rightarrow 0$.
Because thermal noise falls outside the scope of our review, we do not discuss
this topic further. The interested reader is referred to Refs.\
\cite{Mar96,Ave96}.

\section{Quantum Hall Effect}
\label{q7}

In a strong magnetic field the scattering channels of a two-dimensional
electron gas consist of {\em edge states}. Edge states at opposite edges
propagate in opposite directions. In the absence of scattering from one edge to
the other, each of the scattering channels at the Fermi level is transmitted
with probability $T_{n}=1$. This is the regime of the (integer) quantum Hall
effect. (For reviews, see Refs.\ \cite{C&H,Pra90}.) The conductance
$G=(e^{2}/h)\sum_{n}T_{n}$ shows plateaus at integer multiples of $e^{2}/h$ as
a function of magnetic field. The implication for the shot-noise power
$P\propto\sum_{n}T_{n}(1-T_{n})$ is that it should vanish on the plateaus,
similar to the situation in a quantum point contact, see Sec.\ \ref{q1Eb}.

\begin{figure}
\centerline{\psfig{figure=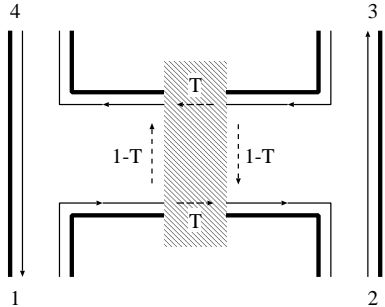,width=5cm}
}%
\caption{
Four-terminal conductor in the regime of the quantum Hall effect. The spatial
location of the edge states is indicated, as well as their direction of motion
(solid lines with arrows). A barrier (shaded region) causes scattering from one
edge to the other (dashed lines). The corresponding scattering probabilities
are indicated.
}\label{QHEfig}
\end{figure}

B\"{u}ttiker \cite{but90,but92b} considered the noise
in the four-terminal conductor depicted in Fig.\ \ref{QHEfig}.
It is assumed that 
the transmission probability of all edge channels 
but one is reduced to zero
by means of a gate across the conductor. 
The remaining non-zero transmission probability is denoted by $T$.
A current flows between contacts 1 and 2 (voltage difference
$V$), while contacts 3 and 4 are voltage probes. This four-terminal
configuration requires a generalization of the two-terminal formulas of Sec.\
\ref{q1D}. The current $I_{a}$ in contact $a$ is related to the
voltages $V_{b}$ at the contact $b$ by the scattering matrix \cite{but86},
\begin{equation}
I_{a}=\frac{e^{2}}{h} 
\left[ V_a \, {\rm Tr} \left( 
{\sf 1} - {\sf s}^{\dagger}_{aa}{\sf s}^{\vphantom{\dagger}}_{aa}
\right)
- \sum_{b\neq a} V_{b} \, {\rm Tr}\,{\sf s}^{\dagger}_{ab}
{\sf s}^{\vphantom{\dagger}}_{ab} \right]
\: ,
\label{Ialphabar}
\end{equation}
where we have assumed zero temperature. The correlator
\begin{equation}
P_{ab}=2\int \limits_{-\infty}^{\infty}\!dt\,\langle\Delta
I_{a}(t+t_{0})\Delta I_{b}(t_{0})\rangle\label{Palphabeta}
\end{equation}
of the current fluctuations in contacts $a$ and $b$ is given by
\cite{but90,but92b}
\begin{equation}
P_{ab}=e\frac{e^{2}}{h}{\sum_{c,d}}'(V_{c}-V_{d})
\left( 
{\rm Tr}\,{\sf s}^{\dagger}_{ac}{\sf s}^{\vphantom{\dagger}}_{ad}
{\sf s}^{\dagger}_{bd}{\sf s}^{\vphantom{\dagger}}_{bc}
+
{\rm Tr}\,{\sf s}^{\dagger}_{ad}{\sf s}^{\vphantom{\dagger}}_{ac}
{\sf s}^{\dagger}_{bc}{\sf s}^{\vphantom{\dagger}}_{bd}
\right)
\: .
\label{Palphabeta2}
\end{equation}
The prime in the summation over $c$ and $d$ means a restriction to
terms with $V_{c}>V_{d}$. The two-terminal formula (\ref{q1D.20}) is
recovered if $c,d=a,b$. Application to the four-terminal
geometry of Fig.\ \ref{QHEfig} shows that $P_{ab}$ vanishes if
$a$ or $b$ equals 1 or 3, while  \cite{but90,but92b}
\begin{equation}
P_{22}=P_{44}=-P_{24}=2eV\frac{e^2}{h}T(1-T) \: .
\label{P11P33}
\end{equation}

Equation (\ref{P11P33}) assumes that the voltages on all terminals are fixed,
while the current fluctuates in time.
Usually, one measures voltage fluctuations at fixed currents.
Current and voltage are linearly
related by Eq.\ (\ref{Ialphabar}), which at low frequencies holds both for the
time-average and for the fluctuations. 
The resulting
voltage-noise power measured between contacts 1 and 4 or between
contacts 2 and 3 is zero. The noise power measured
between any other pair of contacts equals $2eV(h/e^{2})(1-T)/T$  
\cite{but90,but92b}. 
The voltage fluctuations diverge as $T\rightarrow 0$ 
with increasing barrier height. Experiments in support of the edge-channel
description of shot noise in the quantum Hall effect have been reported by
Washburn {\em et al.}\  \cite{was91}.

If the magnetic field becomes so strong that only a single edge channel
remains at the Fermi level, one enters the regime of the {\em
fractional\/} quantum Hall effect. Plateaus in the conductance now
occur at $e^{2}/mh$, $m=1,3,5,\ldots$ (and at other odd-denominator
fractions of $e^{2}/h$ as well) \cite{Pra90}. The quasi-particle
excitations of the electron gas have a fractional charge
$e^{\ast}=e/m$. Since the shot noise, in contrast to the conductance,
is sensitive to the charge of the carriers, one might hope to be able
to find evidence for fractionally charged quasi-particles in shot-noise
measurements. The theory has been developed
in Refs.\  \cite{Kan94,Fen95,Cha95}. 
For very low barrier heights,
Poisson noise with a fractional charge $e^{\ast}$ is expected
in the backscattered current $(I_{\rm max} - I_{1})$,
with $I_{\rm max}=(e^*/h) eV$ the current in the absence of a barrier.
This yields for the shot noise
\begin{equation}
P_{22} = 2 e^* (I_{\rm max} - I_{1}) \: .
\label{q7.1}
\end{equation}
For high barriers the usual Poisson noise
$P=2 eI_1$ is recovered. Experiments remain to be done.

\section*{Acknowledgements}

We would like to thank our collaborators in this research:
H. Birk, M. B\"{u}ttiker, J. I. Dijkhuis, H. van Houten, 
F. Liefrink, L. W. Molenkamp,
and C. Sch\"{o}nenberger.
Permission to reproduce experimental results was kindly given by
M. Reznikov and A. H. Steinbach.
Research at Leiden University is supported by the Dutch Science
Foundation NWO/FOM\@.

\bigskip

This work is to be published in "Mesoscopic Electron Transport,"
edited by L. P.  Kouwenhoven, G. Sch\"{o}n, and L. L. Sohn, 
NATO ASI Series E (Kluwer Academic Publishing, Dordrecht).

\end{document}